\documentclass[a4paper,10pt]{article}
\usepackage[utf8x]{inputenc}
\usepackage{graphicx}
\usepackage{amsmath}
\usepackage[flushleft]{threeparttable}

\title{A comparison of spectra of SBSL produced by standard method and by JB mechanism}
\author{Jaroslav Anto\v{s}\footnote{retired from Institute of Experimental Physics, SAS, Kosice, Slovakia}}

\begin{document}

\maketitle

\begin{abstract}
 Recently  new mechanism to create single bubble sonoluminescence \cite{JBmech} was discovered. Main features of new mechanism
is jet and bubble correlated with this jet which undergoes sonoluminescence. This mechanism will be referenced here
as JB mechanism to produce Single Bubble Sonoluminescence (SBSL).\\
\indent JB mechanism 
extends parameter space where SBSL can be observed. When full
theoretical understanding is missing,
natural question is if properties of SBSL are the same  if SBSL is produced by different mechanisms. 
Partially this question was already addressed in original paper \cite{JBmech}. Conclusion was that within resolution
of the apparatus there are no significant differences between SBSL produced by standard or JB mechanism (in terms 
of amplitude of signal, duration of flash).\\
\indent Aim of this paper is to compare  spectra produced by SBSL
 created by standard method\cite{gaitan} and by JB mechanism.

\end{abstract}
\newpage
\tableofcontents

\section{Introduction}
 Sonoluminescence is a phenomenon which for about 100 years \cite{hist} provokes imagination - sound transformed into light
 - and up to now is still not fully theoretically understood.\\
\indent At first it was discovered as a random flashes of light from bubbles in water under influence of ultrasound. Later this kind 
of sonoluminescence was coined as MBSL - Multi Bubble Sonoluminescence.
 It was difficult to create reproducible results with MBSL and limited theoretical understanding of phenomenon was
 attributed to this fact.\\
 \indent Major step forward was discovery of mechanism
  for creation of sonoluminescence produced by single bubble (so called Single Bubble Sonoluminescence or shortly
SBSL) in 1989 \cite{gaitan}. \\
 \indent After  discovery of SBSL  there have been high expectations that this phenomenon will be studied at well defined
conditions and  theoretically understood soon. SBSL is really excellent object for study and during this
study many unexpected results have been observed. Discovery produced more questions than answers. There are many 
excellent reviews on this topic (e.g. \cite{brenn}). To mention couple surprises - e.g. very short and powerful flash of light.
 About million of photons are radiated from a bubble in about a 50 pico seconds. This radiation is repeated with frequency
$\approx$ 30 kHz (it depends on details of resonator in which SBSL is produced) with clock like precision. Spectrum of light is continuous
with maximum in ultraviolet region and it resembles black-body radiation spectrum. If interpreted as black-body radiation spectrum temperature will correspond 
to  5000  - 20000 \textdegree K. Soon after these findings there appeared theoretical speculations that at right conditions
 temperatures necessary for nuclear fusion can be achieved \cite{fusionT} and later there was claim \cite{taler} that
nuclear fusion was by above mechanism achieved. Significant part of community questions this result.\\
\indent Full theoretical generally accepted understanding is still missing. There is a general agreement about dynamics of bubble
which undergoes sonoluminiscence. Dynamics is described by some versions of Relaygh-Plesset equation \cite{RP}.
 Missing is explanation of properties of flash of light and mechanism of 
creation light by sonoluminescence. \\
 \indent Up to now there are three methods to produce single bubble sonoluminescence. Standard one  is based on 
creating of standing ultrasound wave in resonator filled with proper liquid (e.g. water) with drastically reduced
dissolved gas. At right conditions bubble at position of anti-node periodically shrinks and expands and produces SBSL. 
 This way created SBSL is stable in spatial position and time (in terms of minutes to hours). It is excellent 
object for detailed study.\\
\indent Another method is based on water hammer mechanism \cite{waterham1,waterham2}. Virtue of this method is that it produces sonoluminescence in
bubbles of much larger radius than in case of standard method (400 $\mu$m to be compared with 5 $\mu$m) and up to
about 4 order of larger intensity of light\footnote{Actually this statement depends on working liquid use. If one
compares results for the same liquid difference is not so striking, if any.}.   
Single bubble producing flash of light based on this method is periodically repeated for about 
 10 seconds to minutes and it depends on the 
used working liquid and gas. Liquid in water-hammer pipe should be degassed as in case of standard method.\\
\indent   Recently was proposed new method to produce SBSL \cite{JBmech}. Resonator in new method is exactly the same as in case of
standard method. Important addition is a rod\footnote{in current study rod was composed of copper of diameter 1.7 mm} in the center of the neck of beaver bank. This rod can be moved
 and by a combination of frequency and amplitude of signal generator and position of rod is created jet of 
bubbles which stops at head of jet. At a distance from a head of jet there will be single bubble which interacts 
with jet. This bubble will be producing SBSL.\\
 Stability of position of bubble producing SBSL is not as good as in case of standard method but stable conditions
for generating SBSL (once parameters are set) hold for minutes to hours - as in case of standard method. 
That makes SBSL produced by JB method excellent object of study comparable to one produced by standard method.
 To create jet of bubbles it is necessary to have reasonable level of gas dissolved in liquid. Up to 
saturated level. This requirement is complementary to requirement by standard method or water-hammer method 
 where crucial condition is  drastic reduction of dissolved gas in liquid.
 JB method was tested so far by using only as a working liquid water and as a working gas air. \\
 All three methods represent different approach to create SBSL. One cannot a priory expect that properties of
SBSL produced by different methods are the same. But so far results demonstrate that there are differences
in details  but basic properties (short burst of light, continuum spectrum of light ) 
are the same. \\
 Need for degassing working liquid is not just nuisance which can be neglected. All results about properties
of sonoluminescence are affected by degassing. Experimental data about what properties of working liquid influence
sonoluminescence are quite confusing (see e.g. \cite{waterham2}). Only method to produce SBSL which does not need
degassing of working liquid is JB method. It deserves to be used on more extensive study of the sonoluminescence. \\

\section{Apparatus and SBSL creation}
\begin{figure}
 \includegraphics[width=8 cm]{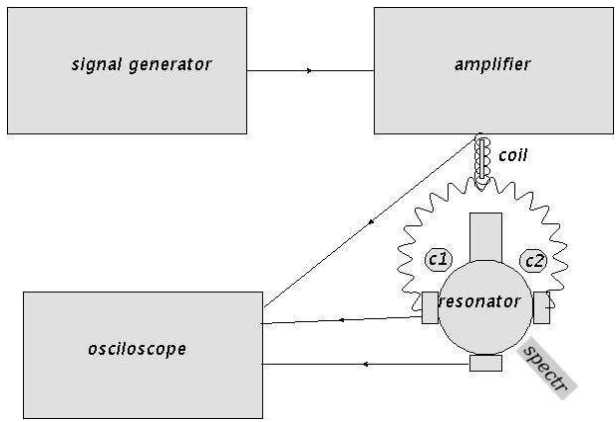}

 \caption{Schema  of our current apparatus  for a study of single bubble sonoluminescence. Resonator is equipped with piezoelements
on sides and microphone (small piezoelement)  bottom,
 C1,2 cameras, spectr - optical system connected to spectrometer }
  \label{schema}
\end{figure}

  Apparatus in this study is the same as in  paper \cite{JBmech}. With one exception that photomultiplier 
is missing (in this setup was not considered necessary) and spectrometer with corresponding optics is included.
 Schematic view of apparatus is in Fig. \ref{schema}. \\
 Resonator is a crucial  for creation of SBSL standard way or by JB method and for this reason is described 
here in more details.
 \subsection{Resonator}
\begin{figure}
 \includegraphics[width=8 cm]{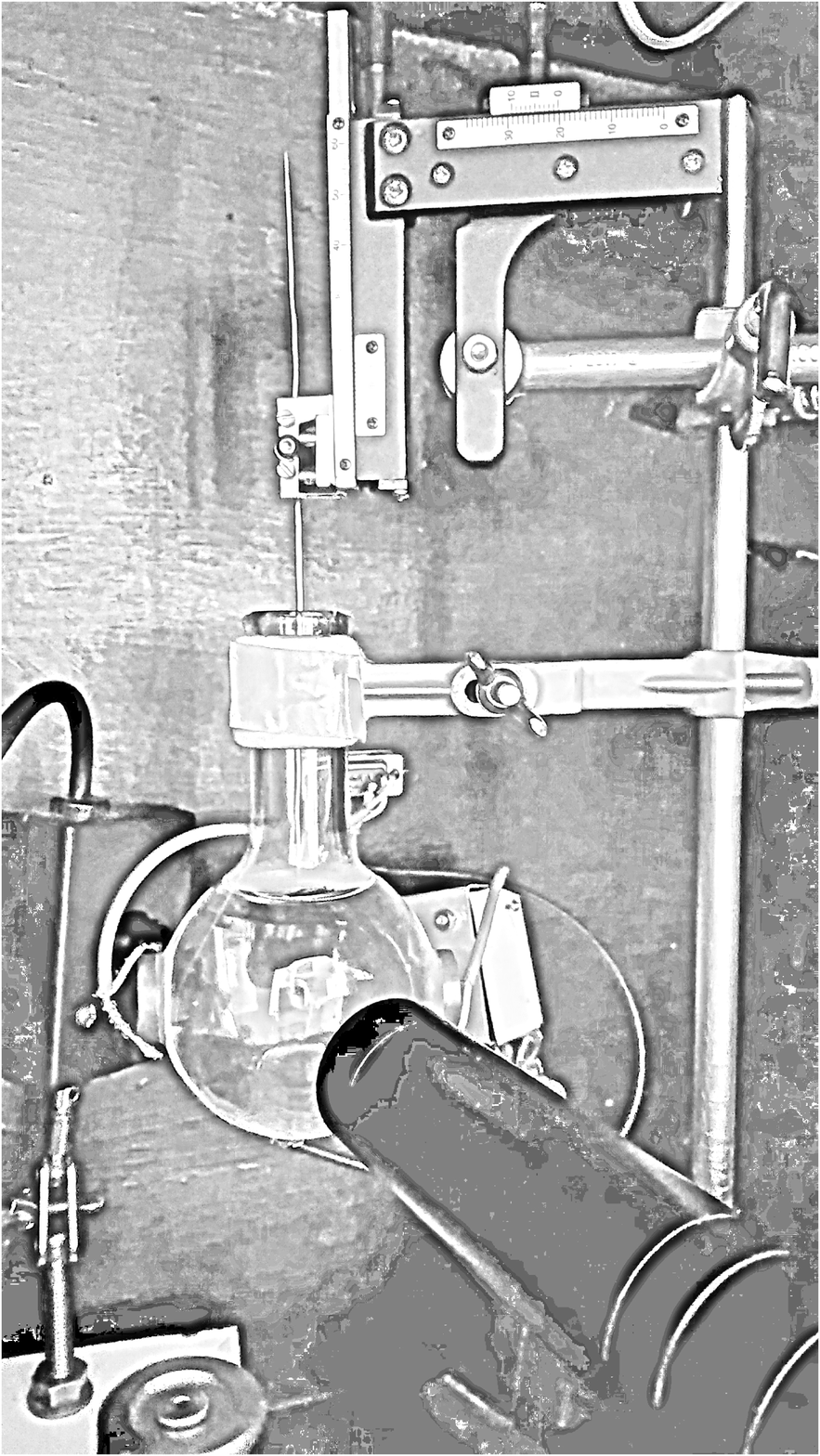}

 \caption{Picture of traditional resonator for study of single bubble sonoluminescence (with rod positioning system included). There are two piezoelement
drivers glued opposite to each other at equatorial side of laboratory 100 ml round bottom flask. Their purpose is to produce
standing ultrasonic wave which will capture bubble at center. At the bottom there is small piezoelement (glued) which serves as microphone. }
  \label{res}
\end{figure}

 Traditional resonator for a study of  SBSL looks like in Fig. \ref{res}. Standard 100 ml beaver boiling 
flask is made from borosilicate glass which strongly suppresses transmission of light for wavelength below 300 nm.\\
 On sides one can see 2 piezoelements. Their purpose is
 to create standing ultrasonic wave in resonator filled by some liquid with small bubble at center. Bubble expands and shrinks in field of
 ultrasonic standing wave and at right conditions (amplitude and frequency of wave) it starts to periodically  burst light - phenomenon known as 
 single bubble sonoluminescence. At bottom of the resonator there is a small piezoelement whose purpose is to detect response of the system.
 It works as a microphone and among other things it helps a great deal to tune system to achieve sonoluminescence. \\
 \indent Traditional way to achieve SBSL using above resonator needs high level of reduction of dissolved air (gas) in working
liquid. Ultrasonic standing wave attracts dissolved gas to the center of resonator. If level of dissolved air is too high
ultrasonic field will rip out dissolved air, form a bubble which is attracted to the center where already bubble is present and interfere with it.
Because of the flow and interaction of attracted bubbles normal regime of expansion and implosion of bubble at center of resonator
leading to SBSL is spoiled. \\
 \indent In a case of very low level of dissolved air in liquid, bubble at center which is supposed to produce SBSL tends 
to dissolve. At this conditions to achieve stable SBSL  is a challenge. \\
 \indent For JB mechanism to work, important addition to resonator is a rod attached to a positioning system (see Fig. \ref{res}). In case of
gas saturated (or close to) liquid at proper conditions there is created jet of bubbles at a distance from a 
single bubble which produces SBSL. Bubbles ripped out of dissolved air are concentrated into jet.

\subsection{Creation of SBSL by JB method}
\begin{figure}
 \includegraphics[width=12 cm]{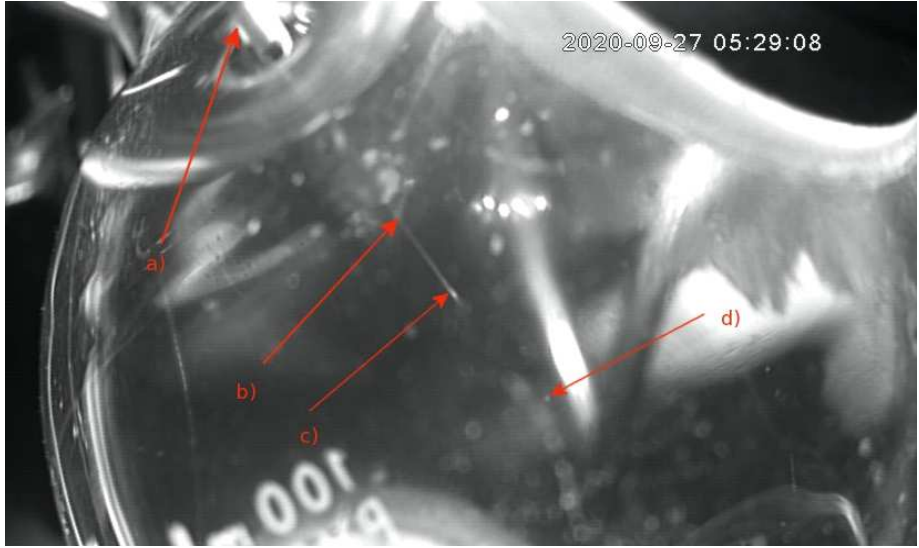}

 \caption{Ingredients of JB method to produce SBSL. In picture there is relevant part of illuminated resonator (tilted by about 45 \textdegree)
with crucial items marked.
a)Points to  part of a narrow cylinder submerged in water. In ultrasonic field it initiate jet (presumably consisting of small
bubbles) to which points arrow marked as b).  c) points to head of jet. Arrow marked as d) points to small dot which is a bubble 
which at above conditions emit light - sonoluminescence. }
\label{nmech}
\end{figure}

 Description of method presented in \cite{JBmech} is repeated here and some more details are added.
 As was already mentioned water (liquid) should not be degassed (contrary to the standard method). A good starting 
point to create SBSL is a same frequency as a one at which SBSL standard way can be created 
(using the same resonator). 
Rod is moved toward a center (see Fig. \ref{schema}) and at some point jet of bubbles toward center of 
resonator is created with distinct head of jet visible. By a combination of moving rod back, modifying frequency and/or amplitude of signal generator
 it is achieved status as in Fig. \ref{nmech}. Close to head of jet, close to a center of resonator, 
is created bubble, which produces sonoluminescence. There is observed occasional interaction between head of jet and a
bubble producing sonoluminescence. In youtube channel \cite{newmech-videos} there are two videoclips
which demonstrate SBSL produced by JB method.  \\

\newpage
\subsection{Spectrometer and optical interface}
 For measurements of spectra of light from SBSL was used spectrometer QE65000 \cite{ocean} connected by optical 
fiber to optical system consisted of 3 lenses made from fused silica with focal legth $f_{l}$=38, -50 and 10 mm respectively.
More details can be found  in paper \cite{spect}. Practically the same setup has been used here. Conditions for 
measurement of spectra have been set as 
follows: integration time 60 s and
 TE cooling applied. At these conditions QE65000 spectrometer has very stable response and dark current is constant.

\section{Measurement of the spectra}
 Main feature of spectra of SBSL produced by standard method is continuous spectrum \cite{spec}, resembling black body radiation
spectrum of temperature 5000-20000 \textdegree K \footnote{there have been recorded spectra which would correspond to
temperature $\approx$ 1000000 \textdegree K \cite{camara}}. In case MBSL (multi bubble sonoluminescence) spectrum is also continuous but
there are also emission lines of constituents \cite{mbsl}. Difference in spectra of SBSL and MBSL are usually
attributed to lower temperature at which radiate bubbles in case of MBSL in comparison with SBSL \footnote{under special
 conditions (working liquid, gas) spectra from SBSL are contain emission lines of constituents see \cite{specOH}-\cite{halide}.}. \\
  Purpose of the current measurement of spectra is to make direct  comparison of spectra of SBSL created by standard
 method and by JB one by
using the same apparatus.  Presented are raw spectra dark current subtracted.

\subsection{Standard method}
\begin{figure*}
\includegraphics[width=10 cm]{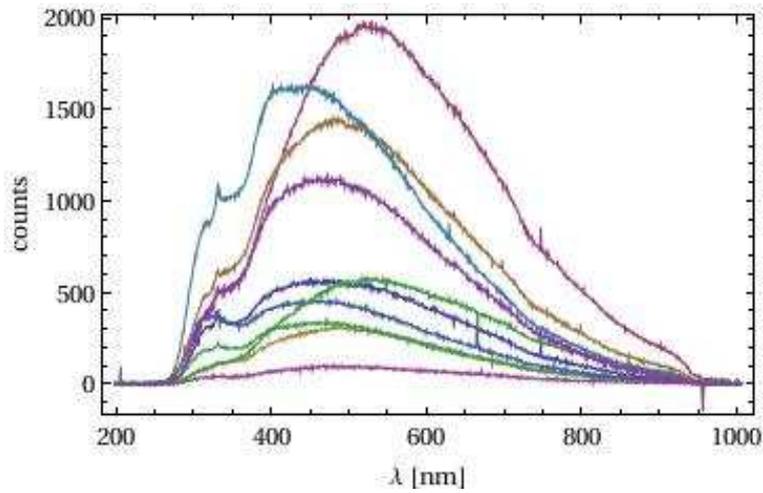}
\caption{Spectra of SBSL. SBSL is created by standard method. }
\label{STNtw}
\end{figure*}

 In Fig. \ref{STNtw} there is a sample of raw spectra of SBSL produced by a standard method. It should be mentioned
that extensive study of spectra of SBSL using different set of resonators (current one among them) was executed
couple years ago \cite{spect} and current results are consistent with previous study. \\
 Current measurements were made at following conditions:\\
Distilled water was degassed to a dissolved oxygen level 1-2 mg/l. SBSL was observed at frequency of
 signal generator f=27.27-27.29 kHz and amplitude (peak to peak) 5-9 $V_{pp}$.\\
 
\newpage
\subsection{JB method}
\begin{figure*}
$\begin{array}{cc}
\includegraphics[width=6 cm]{STspecanaMax.eps} &
 \includegraphics[width=6 cm]{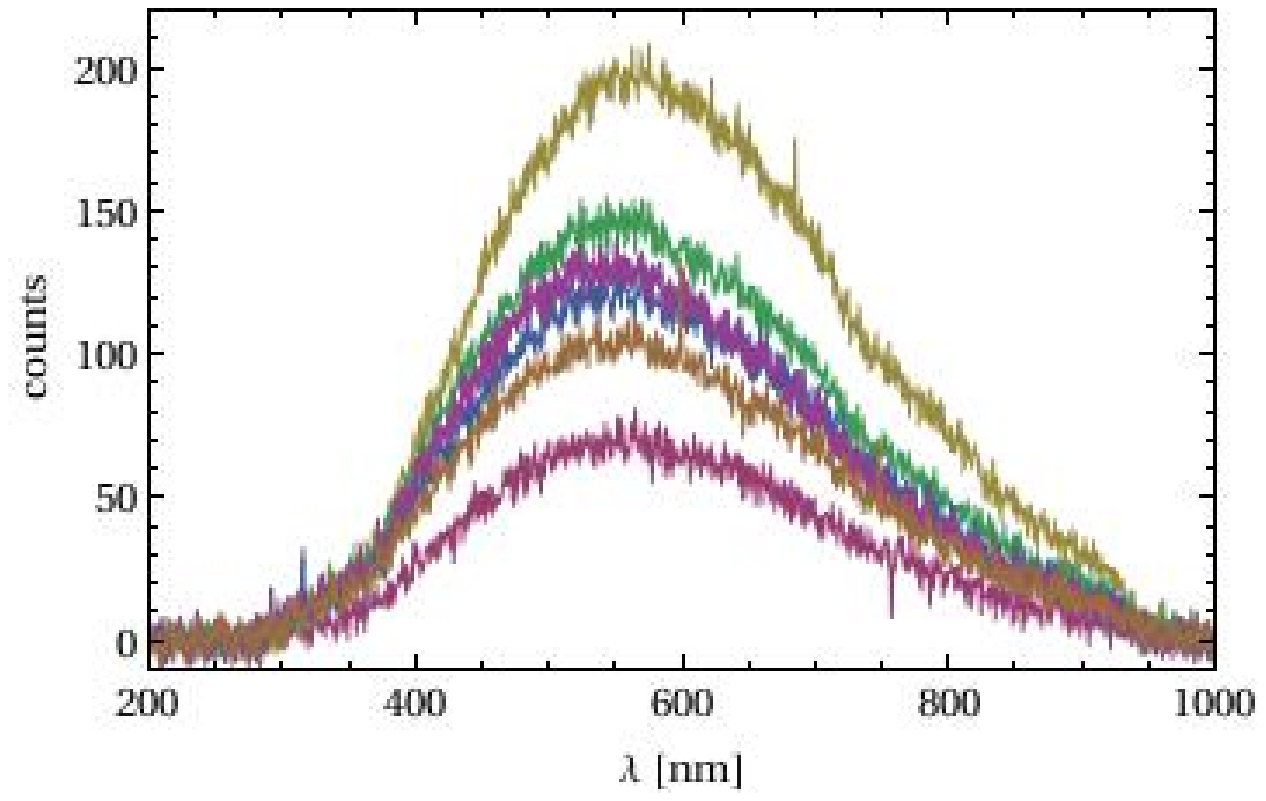} \\
\includegraphics[width=6 cm]{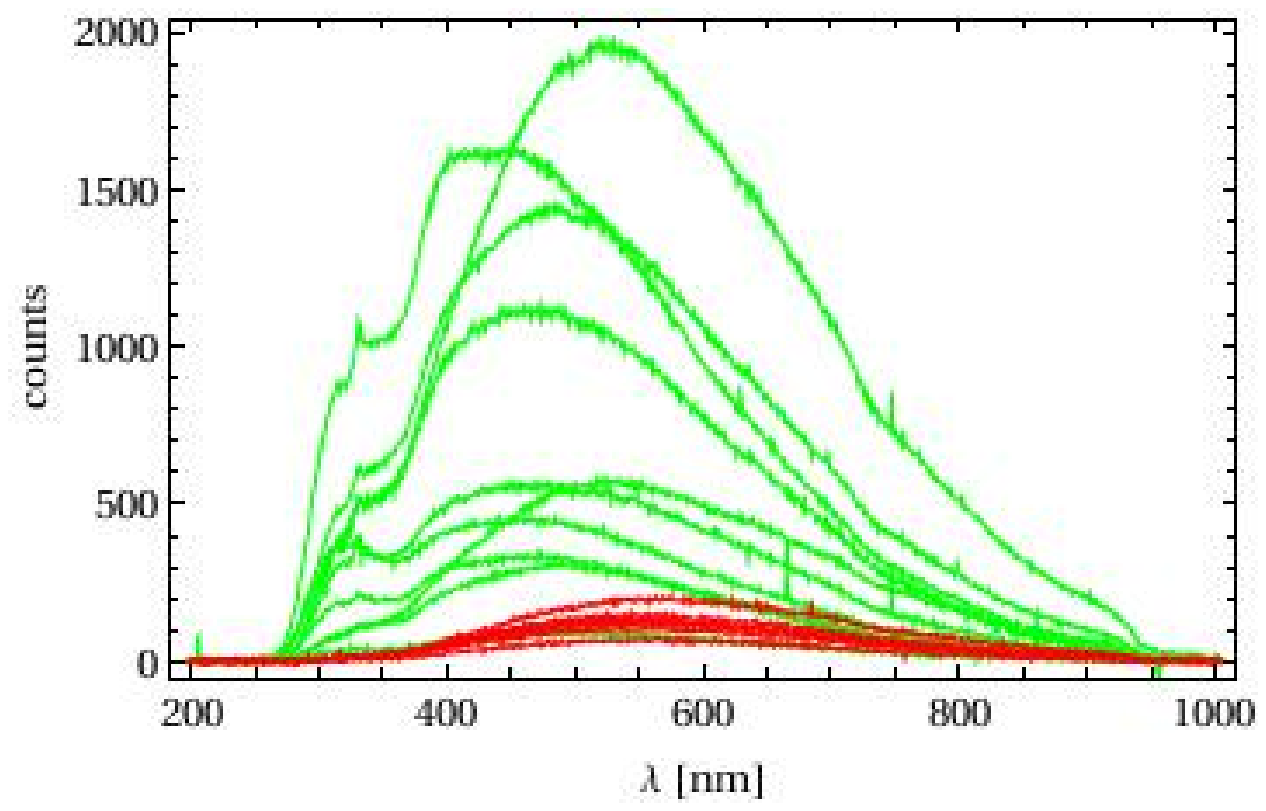} \\
\end{array}$
 \caption{Selected spectra of SBSL produced by standard method (left) and JB method (right). At the bottom there is a comparison
spectra produced standard way (green) and by JB method (red).}
  \label{JBandST}
\end{figure*}

 Working liquid - distilled water was air saturated. SBSL was produced at range of frequencies  27.27 kHz-27.4 kHz and
amplitude       9-10 $V_{pp}$.\\
In Fig. \ref{JBandST} there is a comparison of spectra produced by standard method (upper left) and JB method (upper right). Selected portion is representative one.
 At bottom there is a combined plot where spectra produced by standard method are green and by JB method are red.
 There are couple of obvious differences. Intensity of spectra produced by standard method are order of magnitude
larger. In comparison with JB produced spectra are shifted to lower wavelength. Spectra produced by JB method look
more uniform (even when they have been produced over much larger spread of frequencies. 

\section{Conclusion}

 SBSL spectra produced by JB method share the same characteristics with the standard SBSL spectra. They are continuous, 
 sharp lines related to some constituents (of gas or liquid)  are absent. Spectrometer was wavelength calibrated but not intensity calibrated. Compared are 
just raw dark current subtracted spectra.\\
One can see that spectra created by standard method are shifted to lower wavelength. Interpretation based on 
black body radiation model would conclude that SBSL crated by standard method correspond to higher temperature in
comparison with SBSL created by JB method.  Important condition in 
selection of sample to measure spectra is in both cases stability of SBSL on order of magnitude $\approx$ one hour.
 In previous paper \cite{JBmech} it was claimed that intensity of light produced by JB method and standard method are close.
 But it was also claimed that in case of JB method amplitude of signal is not as uniform as in case of standard method and
for some periods are even missing. That's probably reason why we see order of magnitude greater in case of standard SBSL 
signal when we integrate signal  over 60 seconds. Time dependence of SBSL created by JB method should be studied in more
 details for definite conclusion.\\
 One cannot exclude that by JB method can be obtained spectra of SBSL corresponding to higher temperatures than
by standard method. Because we did not scanned all parameter space available for JB method.  \\
 JB method to produce SBSL is new one and it's potential was not explored yet. There is broad range of options
in a combination of working liquids and gasses which can be explored. There is also some space for
optimization of method itself. Some optimization can be addressed e.g to a rod which initiates jet. More
 sophisticated shape of rod (cone at end of rod ?) and material can prove to be more appropriate for 
specific applications. 
 Detailed study of duration of flash of SBSL produced by JB method (as e.g.  \cite{nature},\cite{timeres}) could be
interesting target for a study. However it is beyond reach of current apparatus.
 \section{Acknowledgments}
 Support from Institute of Experimental Physics by providing spectrometer QE65000 is acknowledged.
Author is gratefull to Professor G. Martinska for carefull reading of manuscript, comments and suggestions.

\end{document}